\documentclass[aps,prstab,twocolumn,groupedaddress,amsmath,nobibnotes,nofootinbib]{revtex4-1}

\usepackage{graphicx}

\usepackage{subcaption}
\usepackage{tabularx}
\usepackage{threeparttablex}
\usepackage{natbib}

\newcommand\Tstrut{\rule{0pt}{2.6ex}}         

\begin{document}

\title{Rapid charge redistribution leading to core hollowing in a high-intensity ion beam}


\author{K. Ruisard, A. Aleksandrov}
\email[]{ruisardkj@ornl.gov}
\affiliation{\textsuperscript{1}Oak Ridge National Laboratory, Oak Ridge, Tennessee 37830, USA}

\date{\today}

\begin{abstract}
Recently, the first direct measurement of a full 6D accelerator beam distribution was reported \cite{Cathey2018}. 
That work observed a correlation between energy and transverse coordinates, for which the energy distribution becomes hollowed and double-peaked near the transverse core.
In this article, a similar structure is shown to emerge in expansion of an initially uncorrelated, high density bunched beam as the result of velocity perturbation from nonlinear space charge forces.
This hollowing is obscured when the 6D phase space is projected onto one- and two-dimensional axes.
This phenomenon has not been widely recognized in accelerator systems, but parallels can be drawn to observations of laser-ionized nanoclusters and electron sources for diffraction. 
While this effect provides insight into the origin of the measured core correlation, it does not provide a complete description.
A better reproduction of the measured structure can be obtained via self-consistent simulation through the radio-frequency quadrupole.

\end{abstract}

\maketitle

\section{Introduction}

Understanding beam dynamics in the early stages of capture and acceleration is crucial in high-intensity accelerators. 
Space charge dynamics in low- to medium-energy transport are suspected to initiate halo formation \cite{Gluckstern1994,Fedotov2003b}, causing uncontrolled losses at higher-power stages.
Improved loss mitigation requires predictive capability of accelerator models.
At present, no model has been shown to deliver accurate representation of the loss-prone beam tails and halo.  
This shortcoming has been attributed to incomplete information of the initial distribution \cite{Qiang2002}, motivating work at the Spallation Neutron Source (SNS) Beam Test Facility (BTF) on complete characterization of the 6D beam phase space \cite{Cathey2018}.

The measurements in \cite{Cathey2018} revealed an interplane correlation that appears as a hollowed, bimodal energy distribution in particle populations near the beam core. 
This feature, which was seen to scale with beam current, was characterized as new and unexpected. 
However, as this article will discuss, a similar phenomenon is broadly observed in a variety of charged particle systems.
It will be shown that conditions leading to core hollowing in these systems can be present in a high-intensity accelerator front-end such as the BTF. 

Work in \cite{Cathey2018} also showed that the energy-hollowing could be reproduced via particle-in-cell (PIC) simulations. 
This article extends that simulation work by taking a closer look at the emergence of this feature.
An initially uncorrelated compact bunch at the exit plane of the radio-frequency quadrupole (RFQ) is seeded and transported through a model of the BTF medium-energy beam transport (MEBT).
Very quickly the bunch develops a transverse-energy dependence resembling observations.
It will be shown that the energy splitting is accompanied by rearrangement of charge into a hollowed distribution, in which excess density ``piles up" near the edge as a result of an outwardly-propagating density perturbation. 
This perturbation is launched by nonlinear space charge in the initial distribution which causes the core to expand more rapidly than the edge.

The key to observing core hollowing is examining slices rather than full projections of the 6D phase space. 
As observed in \cite{Cathey2018}, fully projected distributions may be convex, with a smooth and monotonically decreasing density profile, but simultaneously the profile of a core slice may be flat-topped or hollowed. 
As a result, reliance on standard 2D phase space visualizations may neglect significant core features, particularly for high-charge bunches.  
The slice-emittance approach has not previously been adopted for characterization of ion beams. 
However, this approach is already established in the FEL community, as variation of transverse phase space along the bunch can reduce brightness of the emitted coherent radiation \cite{Dowell2003,Dattoli2012}.

As the demand for high current at high reliability grows, the details of internal, space charge driven core structure will be important for accurately modeling downstream beam dynamics and losses.
One risk of remaining blind to internal core structure is the over-simplification of beam distributions; e.g., adopting a 6D Gaussian on the basis that 1D and 2D projections appear Gaussian. 
It will be shown that the bunch produced by self-consistent tracking of particles through RFQ fields is distinctly non-Gaussian, containing significant internal core structure in qualitative agreement with observations. This article will argue that the pre-existence of internal structure precludes the rapid hollowing from occuring in the MEBT. 

The paper is organized as follows. 
Section \ref{sec:meas} summarizes the relevant observations at the SNS BTF.
Section \ref{sec:lit-review} describes a mechanism for space charge driven hollowing observed in mainly non-accelerator charged particle systems.
Then, Section \ref{sec:sim} examines particle-in-cell simulation of an initial Gaussian distribution in a model of the BTF MEBT. 
A beam with excessively high charge density is used to visualize this mechanism. 
Section \ref{sec:relevance} compares this simple model to output from RFQ simulations.   
Finally, Section \ref{sec:discussion} addresses the implications to simulation and measurement of high-intensity bunched beams.

\begin{figure}[t!]
\includegraphics[width=0.47 \textwidth]{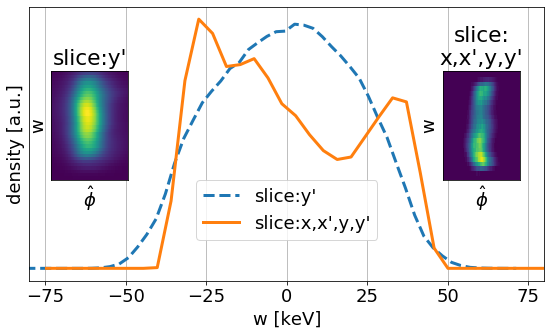}%
\caption{\label{fig:5Dmeas} Measured dependence of energy distribution on dimensionality of a core slice at 21 mA. The longitudinal phase space is plotted in a shear-corrected plane, where $\hat \phi = \phi - \left<\frac{d\phi}{d w}\right> w$.}
\end{figure}

\section{Observations at the SNS Beam Test Facility} \label{sec:meas}


In prior work \cite{Cathey2018}, direct measurement of the 6D distribution was made downstream of the RFQ exit, where a second-order correlation between the energy distribution and all transverse coordinates $(x,x',y,y')$ was observed. 
For transverse coordinates near the beam core, the energy distribution is hollowed. 
For coordinates near the transverse edge and for the fully projected energy distribution, the energy profile is convex and single-peaked.

The importance of studying slices rather than full projections was a key insight adopted during these measurements. 
In order to discuss internal structure of the 6D distribution, it is necessary to distinguish between \textit{full projections}, where information from all particles is projected into a 1D or 2D profile, and \textit{partial projections,} which represent only a fraction of particles selected by slices in the hidden (non-plotted) dimensions.
In the nomenclature adopted here, slice:x,x' indicates selection of slices in the x and x' coordinates. Unless otherwise noted, slice:x,x' bisects the beam core, selecting particles near x=0 and x'=0. The width of the slice is determined by the physical width of slit apertures used in the measurement: $\Delta x  \sim 0.2$ mm and  $\Delta x' \sim 0.3$ mrad.

\begin{figure}[t!]
\includegraphics[width=0.3\textwidth]{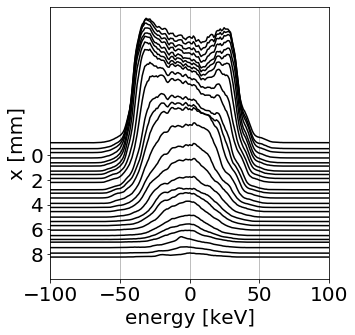}%
\caption{\label{fig:4Dslice} Measured dependence of energy distribution  of a 4D transverse slice on transverse position. Slice:x,x',y,y' is centered over the core in x',y and y'. The center of the slice is varied along x. The beam current is 25 mA.}
\end{figure} 


Figure \ref{fig:5Dmeas} illustrates the observed structure.
The energy distribution of a 1D core slice is compared against a 4D core slice, illustrating hollowing that is hidden in full projections (or, in this case, even in a 1D partial projection).
The dimensionality of a core slice corresponds with proximity of the selected particles to the core, with a 4D core slice containing exclusively low-amplitude core particles. 
The dependence of this hollowing on transverse distance from the core is shown in Figure \ref{fig:4Dslice}. 
This structure was also observed to depend on space charge and is most pronounced at the highest current of 40 mA.

\section{Self-field-driven core hollowing in charged particle bunches} \label{sec:lit-review}

Although not widely seen in accelerator beams, this phenomenon has been observed across diverse fields. 
In the field of laser-plasma interactions, laser ionization of a gaseous nanocluster results in Coulumb explosion (CE). A uniformly filled sphere is an equilibrium distribution in this expansion; however, even perturbative nonuniformity is shown to result in the formation of density rings on the outer edge of the expanding bunch as long as the density decreases from core to outer edge \cite{Kaplan2003,Murphy2014,Grech2011}. 
The initial non-uniformity results in core ions with higher outward velocity than edge ions.
This velocity eventually reaches a crossover point, where inner particle trajectories catch up to outer particles and the radial velocity profile becomes multi-valued.
Density begins to pile up against a critical spherical surface and eventually forms a shock front at the expanding edge.

\begin{figure*}[t]
\includegraphics[width=\textwidth]{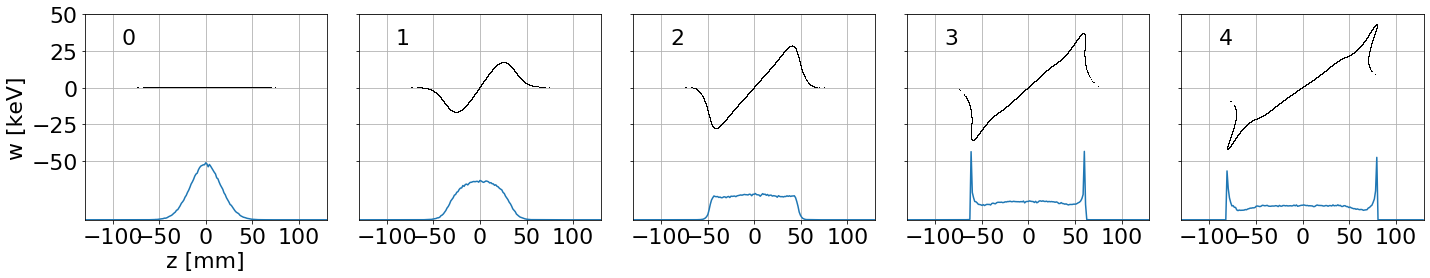}%
\caption{\label{fig:1D} Snapshots of phase space evolution in 1D longitudinal simulation at even intervals of 2.5 meter. A line-out of the spatial profile is also shown.}
\end{figure*}

This effect is also seen in the field of ultrafast electron diffraction (UED), where laser-driven photoemission creates a pancake-shaped bunch ($r>>z$) at the cathode surface. 
In this system, a transverse ring-like density perturbation appears at the expanding edge, resulting in a hollowed transverse profile \cite{Zerbe2018,Zerbe2019}. 
As in the case with the BTF ion beam, the hollow $xy$ profile is only visible for a slice through the longitudinal core, while the fully-projected profile is uniform. 
This study in \cite{Zerbe2018} points to this as one reason this effect was previously undetected. A second reason was the space charge dependence, as the ring only forms at electron densities relevant to next-generation UED sources.

As far as this author is aware, the longitudinal hollowing of a bunched ion beam has not been previously observed.
This may be explained not only by the need to examine high-dimensional phase space slices, but also by the uniqueness of conditions leading to this phenomenon: a compact, high-density expanding bunch.
The transition from RFQ to the MEBT provides such conditions. 

A similar scenario can be found in injectors for bright electron beams. 
Indeed, similar behavior has been noted in modeling of rf photoinjectors \cite{Anderson2000}, where rapid transverse emittance growth was seen to correlate with longitudinal position and was accompanied by a density spike at the outer edge. 
Another parallel can be drawn to the application of laser-wakefield accelerators (LWFA) as electron injectors to colliders or FELs \cite{Chao2003,Gruner2009}, where very small emittance leads to significant space charge effect during bunch expansion between plasma boundary and RF structure. 
Modeling of this system places the same emphasis on knowing the full 6D phase space distribution, and reveals a similar ``deepening" of a bimodal energy distribution with increased space charge \cite{Fubiani2006}.

In two University of Maryland experiments with low-energy, high density unbunched electron beams, the transverse profiles developed visible density rings \cite{Kishek2003,Bernal1999}. 
This study concludes that ``the perturbation is the result of a strong beam-edge imbalance produced by an aperture in an expanding beam" \cite{Kishek2003}.
However, transverse space charge ``rings" may also be indistinguishably driven by external nonlinearity.
Additional structure in \cite{Kishek2003} is shown to originate from velocity perturbation due to field distortion at the cathode grid. 

The systems discussed here share a  mechanism for space charge driven redistribution. 
All cases share the characteristics of non-negligible space charge forces, non-uniformity in initial charge distribution and a freely expanding bunch.
In the following section it will be shown that this same phenomenon is observed in simulations of the BTF MEBT and generates interplane correlations similar to observations (Figures \ref{fig:5Dmeas} and \ref{fig:4Dslice}). 
In the BTF MEBT, the beam initially starts as a tightly focused bunch at the RFQ exit. Within the first meter, the beam expands in all dimensions. Quadrupole focusing limits transverse size, but longitudinally the bunch expands freely.
Therefore, it is consistent with this mechanism that hollowing is observed mainly along the longitudinal dimension.

\section{Hollowing of an initially Gaussian accelerator beam} \label{sec:sim}

Previous work \cite{Cathey2018} demonstrated that the two-peaked energy profile can be recreated through self-consistent particle-in-cell simulation. The same simulation case is revisited here in greater detail. In addition, a 1D longitudinal solver is used for illustration. Both simulations use the PyORBIT code \cite{Shishlo2015}. 
Simulations are done for a 2.5 MeV $H^-$ bunch with $1.55 \times 10^9$ ions per bunch. This is equivalent to 100 mA averaged over an RF cycle, significantly higher than the nominal operating current of 40 mA.
Increasing the space charge drives a more rapid charge redistribution with a deeper and more visible structure. The case at nominal SNS parameters is addressed in Section \ref{sec:relevance}.

First, the 1D longitudinal model is useful for illustrating the effect without high-dimensional slicing.
An initial bunch of 200,000 macroparticles is seeded with a round and uniform transverse distribution of radius 1.3 mm (this is rms equivalent to the profile at the RFQ exit). 
The transverse profile is held constant in the simulation as if subject to uniform transverse focusing.
The initial longitudinal has zero emittance and the spatial distribution is Gaussian with $\sigma_z = 16.5$ mm, about $10\times$ longer than the expected length at the RFQ exit.
In this model, which takes an impedance-based approach to the space charge calculation, a perfectly-conducting cylindrical pipe boundary is assumed at $2\times$ the beam radius. 

Figure \ref{fig:1D} shows the longitudinal evolution in phase space, with the initial distribution shown in frame 0.
In frame 1, a nonlinear velocity kick from the space charge fields is visible and the profile has broadened.
Due to the initial space charge kick, particles with the highest energies are found midway between the core and edge.  
By frame 2, the profile is more uniform. Frame 3 captures a crossover point, where inner particles have caught up to outer particles. At this point the energy distribution $w(z)$ becomes multi-valued and the lower-energy tails start to fold under the expanding core. In the spatial profile, excess density has collected near the edges and two prominent peaks have formed. 
Between frames 3 and 4, the bunch expands with the two peaks ``locked in" at the outer edge.

\begin{table}
\caption{\label{tab:quad} Parameters for quadrupoles in first 1.3 meters of BTF MEBT}
\begin{ruledtabular}
\begin{tabular}{l l l l l}
& s[m] & $\int B \cdot dz $ [T] & $L_{eff} [cm]$ & polarity \\
Quad 1 & 0.131 & 1.149 & 6.1 & $+$\\
Quad 2 & 0.314 & 1.357& 6.6 & $-$\\
Quad 3 & 0.575 & 1.08 & 9.6 & $+$ \\
Quad 4 & 0.771 & 0.61& 9.6 & $-$\\
\end{tabular}
\end{ruledtabular}
\end{table}

\begin{table}
\caption{\label{tab:twiss1} rms parameters for 100 mA Gaussian initial distribution. Emittances are un-normalized.}
\begin{ruledtabular}
\begin{tabular}{llllll}
& x & y & z & \\
$\alpha$      & -2.0 &2.0  &0 &\\ 
$\beta$  [mm/mrad]     &0.2    &0.2  &1.0 (1.3 deg/keV) &\\
$\epsilon$  [mm-mrad] &2.2 & 2.2  & 1.5 (50 deg-keV) &\\
\end{tabular}
\end{ruledtabular}
\end{table}


\begin{figure*}
\centering
\begin{subfigure}[b]{0.3\textwidth}
	\includegraphics[width=\textwidth]{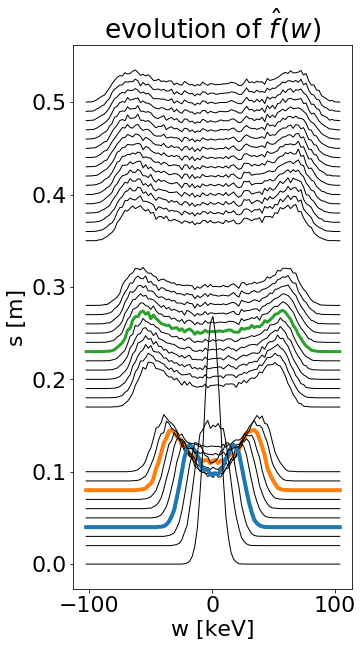}%
	\caption{Evolution of core energy slice}\label{fig:100mAw-waterfall}
\end{subfigure} \hfill
\begin{subfigure}[b]{0.3\textwidth}
	\includegraphics[width=\textwidth]{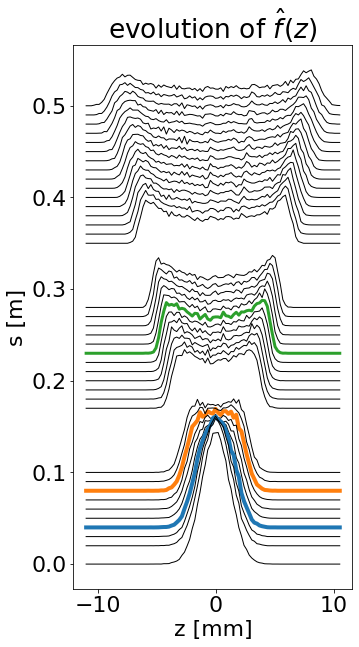}
	\caption{Evolution of core phase slice }\label{fig:100mAz-waterfall}
\end{subfigure} \hfill
\begin{subfigure}[b]{0.33\textwidth}
	\begin{subfigure}[b]{\textwidth}
		\includegraphics[width=\textwidth]{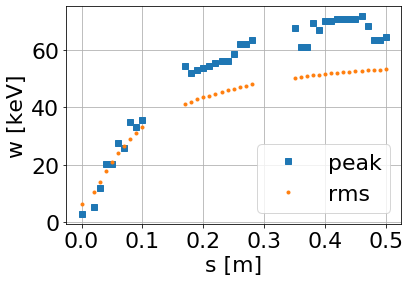}%
		\caption{Evolution of rms and peak energy}\label{fig:100mAw-rms} 
	\end{subfigure} 
	
	\vspace{.1cm}
	\begin{subfigure}[b]{\textwidth}
		\includegraphics[width=\textwidth]{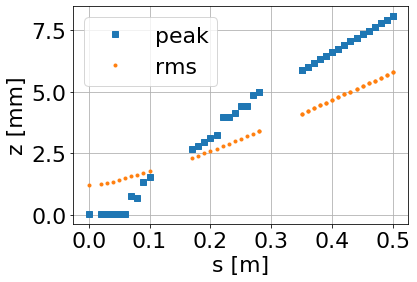}%
		\caption{Evolution of rms and peak phase}\label{fig:100mAz-rms} 
	\end{subfigure}
\end{subfigure}
\caption{\label{fig:100mAevolution} Evolution of the phase and energy profiles of a 10\% core slice. Figures (a) and (b) show the stacked phase and energy profiles for the first 0.5 meters of evolution in the MEBT. The gaps indicate the quadrupole locations. Colored lines highlight the profile at 4 cm, 8 cm and 22 cm. 
Figures (c) and (d) compare the growth rate of the rms width versus the location of the distribution peak. When the distribution is single-peaked it is centered at $z,w = 0$. Once two peaks are formed they propagate outwards and quickly exceed the rms width. Although only one peak is tracked in (c) and (d), the splitting occurs when the peak deviates from 0.}
\end{figure*}

\begin{figure*}
\centering
\begin{subfigure}[t]{0.45\textwidth}
	\includegraphics[width=\textwidth]{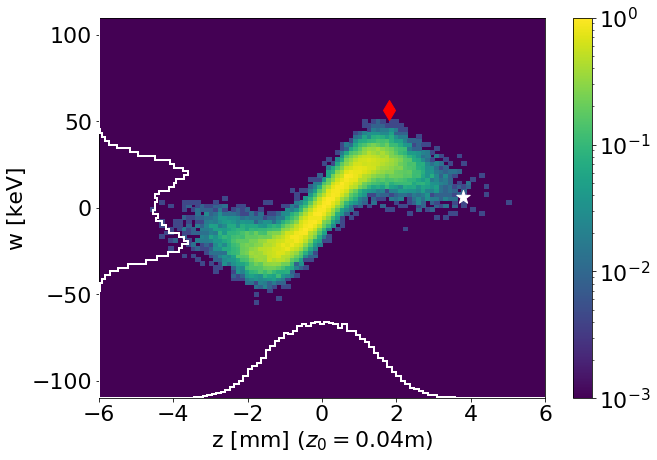}
	\caption{Core-slice projection at $s=4$ cm.}\label{fig:100mAa}
\end{subfigure}
\begin{subfigure}[t]{0.45\textwidth}
	\includegraphics[width=\textwidth]{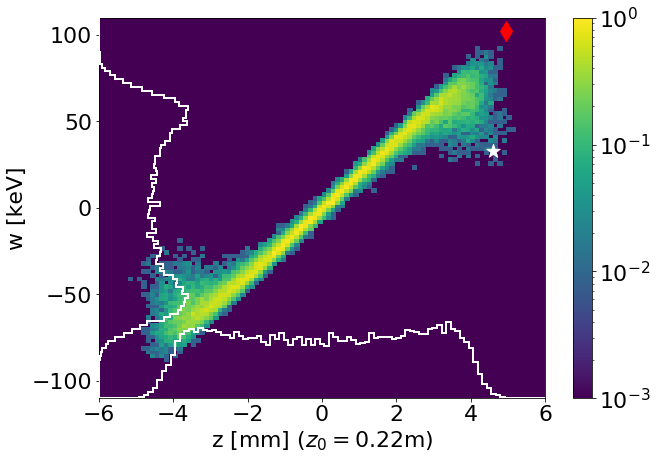}%
	\caption{Core-slice projection at $s=23$ cm.}\label{fig:100mAb}
\end{subfigure}
\begin{subfigure}[t]{0.45\textwidth}
	\includegraphics[width=\textwidth]{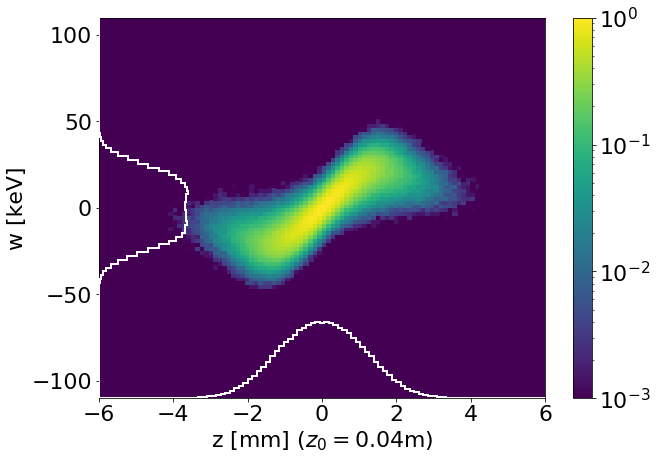}
	\caption{Full projection at $s=4$ cm.}\label{fig:100mAc}
\end{subfigure}
\begin{subfigure}[t]{0.45\textwidth}
	\includegraphics[width=\textwidth]{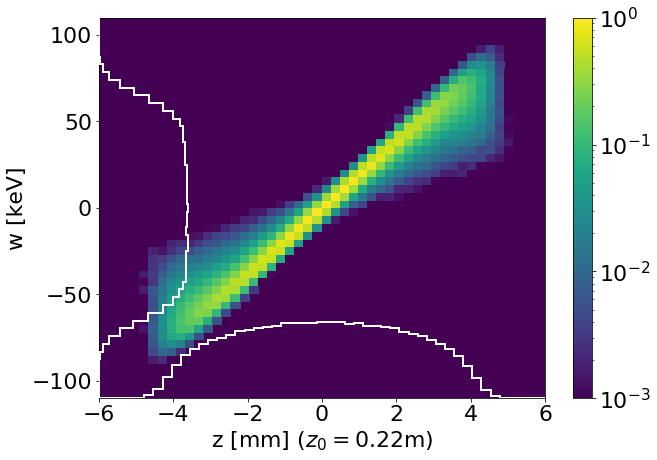}%
	\caption{Full projection at $s=23$ cm.}\label{fig:100mAd}
\end{subfigure}
\caption{\label{fig:100mAframes} Evolution of the 100 mA 3D Gaussian beam. 
Frames (a) and (b) show the projection of the 10\% core slice in $x$ and $y$ at $s=4$ cm and $s=22$ cm. Frames (c) and (d) show the corresponding full projection at both locations. In frames (a) and (b) the location of two individual macro-particles are marked by a red diamond and white star. }
\end{figure*}

\vfill\eject 

The same signatures can be seen in a full 3D simulation of a Gaussian beam when core slicing is applied.
The same case is studied as in \cite{Cathey2018}, which used the code PARMILA to a propagate a bunch through the first meter of the BTF MEBT.
The first meter includes four quadrupoles, which are modeled as hard-edged elements with the parameters reported in Table \ref{tab:quad}.

A Gaussian beam with Twiss parameters shown in Table \ref{tab:twiss1} is seeded at the plane of the RFQ exit. These parameters were chosen to be near the rms parameters expected from the SNS RFQ with two notable differences.
In addition to increasing the current to 100 mA, the longitudinal emittance is half the design value of 100 deg-keV, which further enhances the space charge effect.

The simulation bunch is modeled with 500,000 macroparticles. This is above the requirement for saturation of 100\%  rms parameters, and is chosen to maintain good resolution of the phase space in core slices.
This PyORBIT space charge model uses a 3D FFT Poisson solver with grid size 64x64x64. 
The self fields are calculated every millimeter.

In order to visualize the emergence of core structure, evolution is tracked for a partial distribution containing only particles near the transverse core in $x$ and $y$. 
As the transverse size varies during transport, the slice width is calculated dynamically to contain 10\% of particles in both the $x$ and $y$ 1D projections.

Figure \ref{fig:100mAevolution} depicts the evolution of the core slice phase and energy profiles over the first 50 cm of transport.
Snapshots of the longitudinal phase space at $s=4$ cm and $s=22$ cm are shown in Figure \ref{fig:100mAframes}. 
Within the first few steps, the effects of the space charge force are already apparent as particles accelerate away from the core and the energy distribution broadens. 
At 4 cm, splitting of the energy distribution $w(z)$ is visible (Figure \ref{fig:100mAa}, blue curve in Figure \ref{fig:100mAz-waterfall}). 
The highest-energy particle in this frame is indicated by the red diamond marker. 
Around 8 cm, the energy perturbation has visibly changed the spatial profile (orange curve in Figure \ref{fig:100mAz-waterfall}). It appears broader, and shortly afterwards two distinct peaks appear, framing a hollowed core. This onset is also visible around 8 cm in Figure \ref{fig:100mAz-rms}, where the single peak centered at $z=0$ splits into two peaks moving away from the core. 

The crossover point occurs around 22 cm, when the perturbed inner particles outrun edge particles. This can be seen as the high-energy macro-particle indicated by the red diamond overtakes the edge particle marked by a white star.
The folding-over of the tails in phase space is also visible.
At the point, the edge has reached maximum steepness. 
After this point, indicated by the green curve in Figures \ref{fig:100mAw-waterfall} and \ref{fig:100mAz-waterfall}, the shape of the distribution locks in. The phase profile expands linearly, while the energy width starts to saturate. 

The hollowing of the 100 mA beam is limited to the high current density at the core. For edge particles, the longitudinal distribution remains single-peaked.
As a result, the fully-projected profiles obscure the sharp core features, which is consistent with the observations in \cite{Cathey2018}.
In Figure \ref{fig:100mAc}, the initial velocity perturbation at 4 cm is apparent as a very slight hollowing in the energy profile and barely-visible nonlinear tails in the phase space. 
In Figure \ref{fig:100mAd}, the full projection at 23 cm has a flat-topped energy profile, while the spatial profile is still convex.

\begin{figure}
\includegraphics[width=0.4\textwidth]{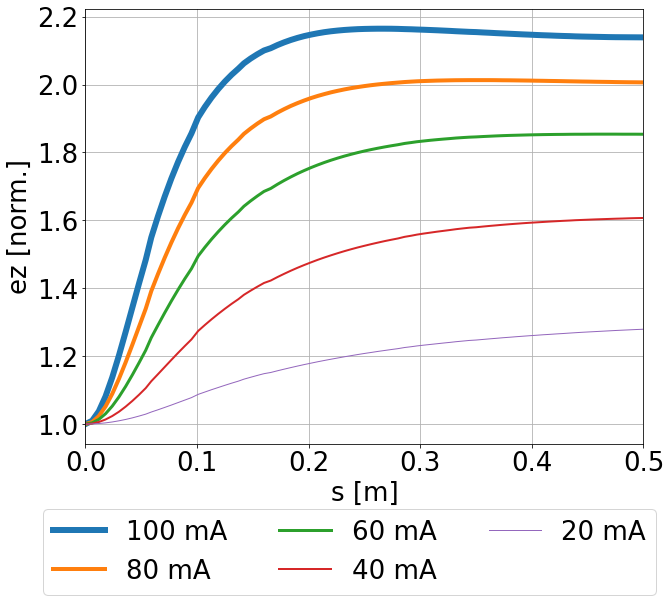}%
\caption{\label{fig:ezgaus} rms emittance evolution for initially Gaussian beam with $\epsilon_{z}=50$ deg-keV, average current as indicated.} 
\end{figure}

Formation of the correlation is accompanied by increase in the longitudinal rms emittance. Unlike the hollowing, this can readily be seen in the full distribution without taking core slices. For the 100 mA case discussed here (Table \ref{tab:twiss1}), the 100\% rms emittance more than doubles in the first 20 cm of transport.
%
Figure \ref{fig:ezgaus} compares emittance growth for different beam currents and fixed rms parameters (using the values in Table \ref{tab:twiss1}).
As charge density decreases, the magnitude of emittance growth decreases and the saturation of emittance growth occurs on a much longer timescale.
As one might expect, in core slices at lower currents the depth of the hollowing decreases and the point at which the phase distribution splits into two peaks occurs later.
In this way, growth of the 100\% emittance can be understood as a signature of the less-visible redistribution of the core density.  


\begin{table}
\caption{\label{tab:twiss2} Output emittances from PARMTEQ simulation of RFQ. Emittances are un-normalized.}
\begin{ruledtabular}
\begin{tabular}{llllll}
output & x & y & z & \Tstrut\\
$\alpha$      & -2.1 &1.6  &0.2 &\\ 
$\beta$  [mm/mrad]     &0.19    &0.14  &0.6 (0.8 deg/keV) &\\
$\epsilon$  [mm-mrad] &3.34 & 3.35  & 3.95 (132 deg-keV) &\\
\end{tabular}
\end{ruledtabular}
\end{table}

\begin{figure*}[th]
\begin{subfigure}[t]{0.4\textwidth}
	\includegraphics[width=\textwidth]{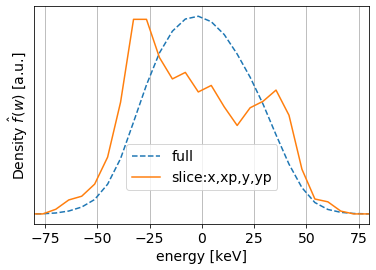}%
	\caption{41 mA RFQ output}\label{fig:wgaussianrfq} 
\end{subfigure}
\begin{subfigure}[t]{0.4\textwidth}
	\includegraphics[width=\textwidth]{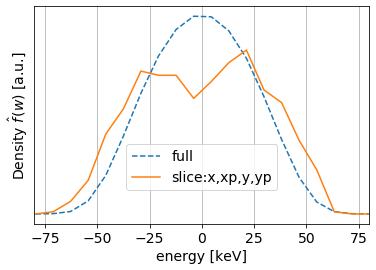}%
	\caption{41 mA Gaussian beam}\label{fig:wgaussian41} 
\end{subfigure}
\caption{Comparison of full and slice energy profiles for simulations with two different initial conditions: (a) 41 mA bunch produced from RFQ simulation and (b) 41 mA Gaussian bunch at RFQ exit. The initial MEBT distributions are rms equivalent.}\label{fig:wcompare}
\end{figure*}

\section{Relevance to measurements}  \label{sec:relevance}

One notable discrepancy between the simulation and measurement is that a much higher average current (100 mA) is needed to drive a visible correlation than was available in measurements (40 mA). 
In addition, the Gaussian distribution expands symmetrically while the observed profile in  Figure \ref{fig:5Dmeas} is consistently lop-sided. 
These discrepancies can be explained by the fact that a Gaussian beam is a poor model for the 6D phase space at the RFQ output. 
A bunch generated by RFQ simulation already contains interplane correlations that also resemble the observed structure.
This is explored in detail in \cite{Ruisard2020a}, but repeated here for illustration. 

The RFQ bunch is generated by PARMTEQ \cite{Crandall1988a} simulation of the 402.5 MHz SNS RFQ \cite{Henderson2014,Ratti2000}. 
The 50 mA input beam has uniform phase, is single-valued in energy at $65$ keV and has a transverse Gaussian distribution. 
5,000,000 macroparticles are used to keep good particle statistics in high-dimensional slices. 
In this case, the slice width is fixed and held equal to the slice width used in 6D phase space measurements.
The output bunch has 41 mA average current with rms Twiss parameters as reported in Table \ref{tab:twiss2}.

Figure \ref{fig:wgaussianrfq} shows the energy profiles of the fully projected bunch and a 4D core slice after propagation in PyORBIT to the measurement plane (1.3 meters downstream of the RFQ). A comparison is made to a Gaussian distribution at the RFQ exit in Figure \ref{fig:wgaussian41}. 
The initial Gaussian bunch is rms-equivalent to the PARMTEQ-generated bunch.
Although the fully projected profile is almost identical between the two cases, the core slice profiles are distinctly different.
The depth and width of the hollowing is very reduced for the 41 mA Gaussian when compared to the 100 mA case described previously. 
In comparison, the RFQ output bunch both has sharper features and qualitatively resembles the measured energy profile in Figure \ref{fig:5Dmeas}. 

Because of the core structure formed in the RFQ, the charge density of the core is already lower and more uniform than in the Gaussian bunch. 
As a result, no significant charge redistribution or emittance growth is observed in propagation through the MEBT.
As the PARMTEQ-generated bunch is understood to be more realistic than the Gaussian bunch, it is unlikely that the mechanism described here is excited in the experiment. 
The hollowing of the bunch occurs earlier within the RFQ, most likely during the initial capture and bunch formation when space charge is most significant.

\section{Discussion}  \label{sec:discussion}

In summary, previous studies at the SNS BTF observed a second-order correlation between energy and transverse coordinates. 
This article describes a simple mechanism by which a Gaussian bunch can rapidly develop a similar structure in medium energy transport. The combination of nonlinear space charge fields and bunch expansion results in hollowing of the core density, an effect which has been seen in other high-density charged particle systems.
While the hollowed core qualitatively resembles observations, self consistent simulation of the RFQ suggests that in the laboratory this structure develops earlier. 

Although insight from this mechanism may be useful in understanding the origin of core-hollowing within the RFQ, the dynamics may be quite different. 
The hollowing described here depends on bunch expansion: as the space charge density drops rapidly, the nonlinearity from the initial distribution is imprinted as a velocity perturbation. In contrast, within the RFQ the bunch is immersed in transverse and longitudinal focusing. Any expansion occurs in the presence of external fields. 

As this phenomenon is not expected to occur in a realistic MEBT bunch, effort should be made to avoid artificially inducing it in simulation.
This study illustrates how a space charge driven correlation may be obscured in typical full projections and reinforces the need to create initial distributions with realistic, fully correlated 6D distributions. 
As demonstrated, spurious emittance growth can be driven when a convex initial distribution is inappropriately assumed. 
To avoid over-estimating the space charge force, extra care must be taken to include the effects of upstream transverse-longitudinal coupling when generating high intensity simulation. 
The implications of this core structure for other space charge driven effects, such as halo formation and beam loss, is not yet understood and will be a focus in future studies.

\subsection{Acknowledgments}
 
The authors are very grateful for assistance with PyORBIT and invaluable feedback on drafts provided by Jeff Holmes, Andrei Shishlo and Sarah Cousineau. 
Thanks also to George Hine, who first recognized the similarity between the BTF observations and the physics of laser-ionized clusters.
This manuscript has been authored by UT-Battelle, LLC, under Contract No. DE-AC0500OR22725 with the U.S. Department of Energy.
This research used resources at the Spallation Neutron Source, a DOE Office of Science User Facility operated by the Oak Ridge National Laboratory.

\bibliography{btf-charge-redistribution}

\begin{thebibliography}{22}%
\makeatletter
\providecommand \@ifxundefined [1]{%
 \@ifx{#1\undefined}
}%
\providecommand \@ifnum [1]{%
 \ifnum #1\expandafter \@firstoftwo
 \else \expandafter \@secondoftwo
 \fi
}%
\providecommand \@ifx [1]{%
 \ifx #1\expandafter \@firstoftwo
 \else \expandafter \@secondoftwo
 \fi
}%
\providecommand \natexlab [1]{#1}%
\providecommand \enquote  [1]{``#1''}%
\providecommand \bibnamefont  [1]{#1}%
\providecommand \bibfnamefont [1]{#1}%
\providecommand \citenamefont [1]{#1}%
\providecommand \href@noop [0]{\@secondoftwo}%
\providecommand \href [0]{\begingroup \@sanitize@url \@href}%
\providecommand \@href[1]{\@@startlink{#1}\@@href}%
\providecommand \@@href[1]{\endgroup#1\@@endlink}%
\providecommand \@sanitize@url [0]{\catcode `\\12\catcode `\$12\catcode
  `\&12\catcode `\#12\catcode `\^12\catcode `\_12\catcode `\%12\relax}%
\providecommand \@@startlink[1]{}%
\providecommand \@@endlink[0]{}%
\providecommand \url  [0]{\begingroup\@sanitize@url \@url }%
\providecommand \@url [1]{\endgroup\@href {#1}{\urlprefix }}%
\providecommand \urlprefix  [0]{URL }%
\providecommand \Eprint [0]{\href }%
\providecommand \doibase [0]{http://dx.doi.org/}%
\providecommand \selectlanguage [0]{\@gobble}%
\providecommand \bibinfo  [0]{\@secondoftwo}%
\providecommand \bibfield  [0]{\@secondoftwo}%
\providecommand \translation [1]{[#1]}%
\providecommand \BibitemOpen [0]{}%
\providecommand \bibitemStop [0]{}%
\providecommand \bibitemNoStop [0]{.\EOS\space}%
\providecommand \EOS [0]{\spacefactor3000\relax}%
\providecommand \BibitemShut  [1]{\csname bibitem#1\endcsname}%
\let\auto@bib@innerbib\@empty
\bibitem [{\citenamefont {Cathey}\ \emph {et~al.}(2018)\citenamefont {Cathey},
  \citenamefont {Cousineau}, \citenamefont {Aleksandrov},\ and\ \citenamefont
  {Zhukov}}]{Cathey2018}%
  \BibitemOpen
  \bibfield  {author} {\bibinfo {author} {\bibfnamefont {B.}~\bibnamefont
  {Cathey}}, \bibinfo {author} {\bibfnamefont {S.}~\bibnamefont {Cousineau}},
  \bibinfo {author} {\bibfnamefont {A.}~\bibnamefont {Aleksandrov}}, \ and\
  \bibinfo {author} {\bibfnamefont {A.}~\bibnamefont {Zhukov}},\ }\href
  {\doibase 10.1103/PhysRevLett.121.064804} {\bibfield  {journal} {\bibinfo
  {journal} {Physical Review Letters}\ }\textbf {\bibinfo {volume} {121}},\
  \bibinfo {pages} {064804} (\bibinfo {year} {2018})}\BibitemShut {NoStop}%
\bibitem [{\citenamefont {Gluckstern}(1994)}]{Gluckstern1994}%
  \BibitemOpen
  \bibfield  {author} {\bibinfo {author} {\bibfnamefont {R.~L.}\ \bibnamefont
  {Gluckstern}},\ }\href@noop {} {\bibfield  {journal} {\bibinfo  {journal}
  {Physical Review Letters}\ }\textbf {\bibinfo {volume} {73}},\ \bibinfo
  {pages} {1247} (\bibinfo {year} {1994})}\BibitemShut {NoStop}%
\bibitem [{\citenamefont {Fedotov}(2003)}]{Fedotov2003b}%
  \BibitemOpen
  \bibfield  {author} {\bibinfo {author} {\bibfnamefont {A.~V.}\ \bibnamefont
  {Fedotov}},\ }\href {\doibase 10.1063/1.1638311} {\bibfield  {journal}
  {\bibinfo  {journal} {AIP Conference Proceedings}\ }\textbf {\bibinfo
  {volume} {3}},\ \bibinfo {pages} {3} (\bibinfo {year} {2003})}\BibitemShut
  {NoStop}%
\bibitem [{\citenamefont {Qiang}\ \emph {et~al.}(2002)\citenamefont {Qiang},
  \citenamefont {Colestock}, \citenamefont {Gilpatrick}, \citenamefont {Smith},
  \citenamefont {Wangler},\ and\ \citenamefont {Schulze}}]{Qiang2002}%
  \BibitemOpen
  \bibfield  {author} {\bibinfo {author} {\bibfnamefont {J.}~\bibnamefont
  {Qiang}}, \bibinfo {author} {\bibfnamefont {P.~L.}\ \bibnamefont
  {Colestock}}, \bibinfo {author} {\bibfnamefont {D.}~\bibnamefont
  {Gilpatrick}}, \bibinfo {author} {\bibfnamefont {H.~V.}\ \bibnamefont
  {Smith}}, \bibinfo {author} {\bibfnamefont {T.~P.}\ \bibnamefont {Wangler}},
  \ and\ \bibinfo {author} {\bibfnamefont {M.~E.}\ \bibnamefont {Schulze}},\
  }\href {\doibase 10.1103/PhysRevSTAB.5.124201} {\bibfield  {journal}
  {\bibinfo  {journal} {Physical Review Special Topics - Accelerators and
  Beams}\ }\textbf {\bibinfo {volume} {5}},\ \bibinfo {pages} {35} (\bibinfo
  {year} {2002})}\BibitemShut {NoStop}%
\bibitem [{\citenamefont {Dowell}\ \emph {et~al.}(2003)\citenamefont {Dowell},
  \citenamefont {Bolton}, \citenamefont {Clendenin}, \citenamefont {Emma},
  \citenamefont {Gierman}, \citenamefont {Graves}, \citenamefont {Limborg},
  \citenamefont {Murphy},\ and\ \citenamefont {Schmerge}}]{Dowell2003}%
  \BibitemOpen
  \bibfield  {author} {\bibinfo {author} {\bibfnamefont {D.~H.}\ \bibnamefont
  {Dowell}}, \bibinfo {author} {\bibfnamefont {P.~R.}\ \bibnamefont {Bolton}},
  \bibinfo {author} {\bibfnamefont {J.~E.}\ \bibnamefont {Clendenin}}, \bibinfo
  {author} {\bibfnamefont {P.}~\bibnamefont {Emma}}, \bibinfo {author}
  {\bibfnamefont {S.~M.}\ \bibnamefont {Gierman}}, \bibinfo {author}
  {\bibfnamefont {W.~S.}\ \bibnamefont {Graves}}, \bibinfo {author}
  {\bibfnamefont {C.~G.}\ \bibnamefont {Limborg}}, \bibinfo {author}
  {\bibfnamefont {B.~F.}\ \bibnamefont {Murphy}}, \ and\ \bibinfo {author}
  {\bibfnamefont {J.~F.}\ \bibnamefont {Schmerge}},\ }\href {\doibase
  10.1016/S0168-9002(03)00939-2} {\bibfield  {journal} {\bibinfo  {journal}
  {Nuclear Instruments and Methods in Physics Research, Section A:
  Accelerators, Spectrometers, Detectors and Associated Equipment}\ }\textbf
  {\bibinfo {volume} {507}},\ \bibinfo {pages} {327} (\bibinfo {year}
  {2003})}\BibitemShut {NoStop}%
\bibitem [{\citenamefont {Dattoli}\ \emph {et~al.}(2012)\citenamefont
  {Dattoli}, \citenamefont {Sabia}, \citenamefont {Ronsivalle}, \citenamefont
  {{Del Franco}},\ and\ \citenamefont {Petralia}}]{Dattoli2012}%
  \BibitemOpen
  \bibfield  {author} {\bibinfo {author} {\bibfnamefont {G.}~\bibnamefont
  {Dattoli}}, \bibinfo {author} {\bibfnamefont {E.}~\bibnamefont {Sabia}},
  \bibinfo {author} {\bibfnamefont {C.}~\bibnamefont {Ronsivalle}}, \bibinfo
  {author} {\bibfnamefont {M.}~\bibnamefont {{Del Franco}}}, \ and\ \bibinfo
  {author} {\bibfnamefont {A.}~\bibnamefont {Petralia}},\ }\href {\doibase
  10.1016/j.nima.2011.12.099} {\bibfield  {journal} {\bibinfo  {journal}
  {Nuclear Instruments and Methods in Physics Research, Section A:
  Accelerators, Spectrometers, Detectors and Associated Equipment}\ }\textbf
  {\bibinfo {volume} {671}},\ \bibinfo {pages} {51} (\bibinfo {year}
  {2012})}\BibitemShut {NoStop}%
\bibitem [{\citenamefont {Kaplan}\ \emph {et~al.}(2003)\citenamefont {Kaplan},
  \citenamefont {Dubetsky},\ and\ \citenamefont {Shkolnikov}}]{Kaplan2003}%
  \BibitemOpen
  \bibfield  {author} {\bibinfo {author} {\bibfnamefont {A.~E.}\ \bibnamefont
  {Kaplan}}, \bibinfo {author} {\bibfnamefont {B.~Y.}\ \bibnamefont
  {Dubetsky}}, \ and\ \bibinfo {author} {\bibfnamefont {P.~L.}\ \bibnamefont
  {Shkolnikov}},\ }\href {\doibase 10.1103/PhysRevLett.91.143401} {\bibfield
  {journal} {\bibinfo  {journal} {Physical Review Letters}\ }\textbf {\bibinfo
  {volume} {91}},\ \bibinfo {pages} {9} (\bibinfo {year} {2003})}\BibitemShut
  {NoStop}%
\bibitem [{\citenamefont {Murphy}\ \emph {et~al.}(2014)\citenamefont {Murphy},
  \citenamefont {Speirs}, \citenamefont {Sheludko}, \citenamefont {Putkunz},
  \citenamefont {McCulloch}, \citenamefont {Sparkes},\ and\ \citenamefont
  {Scholten}}]{Murphy2014}%
  \BibitemOpen
  \bibfield  {author} {\bibinfo {author} {\bibfnamefont {D.}~\bibnamefont
  {Murphy}}, \bibinfo {author} {\bibfnamefont {R.~W.}\ \bibnamefont {Speirs}},
  \bibinfo {author} {\bibfnamefont {D.~V.}\ \bibnamefont {Sheludko}}, \bibinfo
  {author} {\bibfnamefont {C.~T.}\ \bibnamefont {Putkunz}}, \bibinfo {author}
  {\bibfnamefont {A.~J.}\ \bibnamefont {McCulloch}}, \bibinfo {author}
  {\bibfnamefont {B.~M.}\ \bibnamefont {Sparkes}}, \ and\ \bibinfo {author}
  {\bibfnamefont {R.~E.}\ \bibnamefont {Scholten}},\ }\href {\doibase
  10.1038/ncomms5489} {\bibfield  {journal} {\bibinfo  {journal} {Nature
  Communications}\ }\textbf {\bibinfo {volume} {5}},\ \bibinfo {pages} {1}
  (\bibinfo {year} {2014})}\BibitemShut {NoStop}%
\bibitem [{\citenamefont {Grech}\ \emph {et~al.}(2011)\citenamefont {Grech},
  \citenamefont {Nuter}, \citenamefont {Mikaberidze}, \citenamefont {{Di
  Cintio}}, \citenamefont {Gremillet}, \citenamefont {Lefebvre}, \citenamefont
  {Saalmann}, \citenamefont {Rost},\ and\ \citenamefont {Skupin}}]{Grech2011}%
  \BibitemOpen
  \bibfield  {author} {\bibinfo {author} {\bibfnamefont {M.}~\bibnamefont
  {Grech}}, \bibinfo {author} {\bibfnamefont {R.}~\bibnamefont {Nuter}},
  \bibinfo {author} {\bibfnamefont {A.}~\bibnamefont {Mikaberidze}}, \bibinfo
  {author} {\bibfnamefont {P.}~\bibnamefont {{Di Cintio}}}, \bibinfo {author}
  {\bibfnamefont {L.}~\bibnamefont {Gremillet}}, \bibinfo {author}
  {\bibfnamefont {E.}~\bibnamefont {Lefebvre}}, \bibinfo {author}
  {\bibfnamefont {U.}~\bibnamefont {Saalmann}}, \bibinfo {author}
  {\bibfnamefont {J.~M.}\ \bibnamefont {Rost}}, \ and\ \bibinfo {author}
  {\bibfnamefont {S.}~\bibnamefont {Skupin}},\ }\href {\doibase
  10.1103/PhysRevE.84.056404} {\bibfield  {journal} {\bibinfo  {journal}
  {Physical Review E - Statistical, Nonlinear, and Soft Matter Physics}\
  }\textbf {\bibinfo {volume} {84}},\ \bibinfo {pages} {1} (\bibinfo {year}
  {2011})}\BibitemShut {NoStop}%
\bibitem [{\citenamefont {Zerbe}\ \emph {et~al.}(2018)\citenamefont {Zerbe},
  \citenamefont {Xiang}, \citenamefont {Ruan}, \citenamefont {Lund},\ and\
  \citenamefont {Duxbury}}]{Zerbe2018}%
  \BibitemOpen
  \bibfield  {author} {\bibinfo {author} {\bibfnamefont {B.~S.}\ \bibnamefont
  {Zerbe}}, \bibinfo {author} {\bibfnamefont {X.}~\bibnamefont {Xiang}},
  \bibinfo {author} {\bibfnamefont {C.~Y.}\ \bibnamefont {Ruan}}, \bibinfo
  {author} {\bibfnamefont {S.~M.}\ \bibnamefont {Lund}}, \ and\ \bibinfo
  {author} {\bibfnamefont {P.~M.}\ \bibnamefont {Duxbury}},\ }\href {\doibase
  10.1103/PhysRevAccelBeams.21.064201} {\bibfield  {journal} {\bibinfo
  {journal} {Physical Review Accelerators and Beams}\ }\textbf {\bibinfo
  {volume} {21}},\ \bibinfo {pages} {64201} (\bibinfo {year}
  {2018})}\BibitemShut {NoStop}%
\bibitem [{\citenamefont {Zerbe}\ and\ \citenamefont
  {Duxbury}(2019)}]{Zerbe2019}%
  \BibitemOpen
  \bibfield  {author} {\bibinfo {author} {\bibfnamefont {B.~S.}\ \bibnamefont
  {Zerbe}}\ and\ \bibinfo {author} {\bibfnamefont {P.~M.}\ \bibnamefont
  {Duxbury}},\ }\href {\doibase 10.1103/PhysRevAccelBeams.22.114402} {\bibfield
   {journal} {\bibinfo  {journal} {Physical Review Accelerators and Beams}\
  }\textbf {\bibinfo {volume} {22}},\ \bibinfo {pages} {114402} (\bibinfo
  {year} {2019})}\BibitemShut {NoStop}%
\bibitem [{\citenamefont {Anderson}\ and\ \citenamefont
  {Rosenzweig}(2000)}]{Anderson2000}%
  \BibitemOpen
  \bibfield  {author} {\bibinfo {author} {\bibfnamefont {S.~G.}\ \bibnamefont
  {Anderson}}\ and\ \bibinfo {author} {\bibfnamefont {J.~B.}\ \bibnamefont
  {Rosenzweig}},\ }\href {\doibase 10.1142/9789812792181_0016} {\bibfield
  {journal} {\bibinfo  {journal} {Physical Review Special Topics - Accelerators
  and Beams}\ }\textbf {\bibinfo {volume} {3}},\ \bibinfo {pages} {69}
  (\bibinfo {year} {2000})}\BibitemShut {NoStop}%
\bibitem [{\citenamefont {Chao}\ \emph {et~al.}(2003)\citenamefont {Chao},
  \citenamefont {Pitthan}, \citenamefont {Tajima},\ and\ \citenamefont
  {Yeremian}}]{Chao2003}%
  \BibitemOpen
  \bibfield  {author} {\bibinfo {author} {\bibfnamefont {A.~W.}\ \bibnamefont
  {Chao}}, \bibinfo {author} {\bibfnamefont {R.}~\bibnamefont {Pitthan}},
  \bibinfo {author} {\bibfnamefont {T.}~\bibnamefont {Tajima}}, \ and\ \bibinfo
  {author} {\bibfnamefont {D.}~\bibnamefont {Yeremian}},\ }\href {\doibase
  10.1103/PhysRevSTAB.6.024201} {\bibfield  {journal} {\bibinfo  {journal}
  {Physical Review Special Topics - Accelerators and Beams}\ }\textbf {\bibinfo
  {volume} {6}},\ \bibinfo {pages} {19} (\bibinfo {year} {2003})}\BibitemShut
  {NoStop}%
\bibitem [{\citenamefont {Gr{\"{u}}ner}\ \emph {et~al.}(2009)\citenamefont
  {Gr{\"{u}}ner}, \citenamefont {Schroeder}, \citenamefont {Maier},
  \citenamefont {Becker},\ and\ \citenamefont {Mikhailova}}]{Gruner2009}%
  \BibitemOpen
  \bibfield  {author} {\bibinfo {author} {\bibfnamefont {F.~J.}\ \bibnamefont
  {Gr{\"{u}}ner}}, \bibinfo {author} {\bibfnamefont {C.~B.}\ \bibnamefont
  {Schroeder}}, \bibinfo {author} {\bibfnamefont {A.~R.}\ \bibnamefont
  {Maier}}, \bibinfo {author} {\bibfnamefont {S.}~\bibnamefont {Becker}}, \
  and\ \bibinfo {author} {\bibfnamefont {J.~M.}\ \bibnamefont {Mikhailova}},\
  }\href {\doibase 10.1103/PhysRevSTAB.12.020701} {\bibfield  {journal}
  {\bibinfo  {journal} {Physical Review Special Topics - Accelerators and
  Beams}\ }\textbf {\bibinfo {volume} {12}},\ \bibinfo {pages} {1} (\bibinfo
  {year} {2009})}\BibitemShut {NoStop}%
\bibitem [{\citenamefont {Fubiani}\ \emph {et~al.}(2006)\citenamefont
  {Fubiani}, \citenamefont {Qiang}, \citenamefont {Esarey}, \citenamefont
  {Leemans},\ and\ \citenamefont {Dugan}}]{Fubiani2006}%
  \BibitemOpen
  \bibfield  {author} {\bibinfo {author} {\bibfnamefont {G.}~\bibnamefont
  {Fubiani}}, \bibinfo {author} {\bibfnamefont {J.}~\bibnamefont {Qiang}},
  \bibinfo {author} {\bibfnamefont {E.}~\bibnamefont {Esarey}}, \bibinfo
  {author} {\bibfnamefont {W.~P.}\ \bibnamefont {Leemans}}, \ and\ \bibinfo
  {author} {\bibfnamefont {G.}~\bibnamefont {Dugan}},\ }\href {\doibase
  10.1103/PhysRevSTAB.9.064402} {\bibfield  {journal} {\bibinfo  {journal}
  {Physical Review Special Topics - Accelerators and Beams}\ }\textbf {\bibinfo
  {volume} {9}},\ \bibinfo {pages} {064402} (\bibinfo {year}
  {2006})}\BibitemShut {NoStop}%
\bibitem [{\citenamefont {Kishek}\ \emph {et~al.}(2003)\citenamefont {Kishek},
  \citenamefont {Bernal}, \citenamefont {Bohn}, \citenamefont {Grote},
  \citenamefont {Haber}, \citenamefont {Li}, \citenamefont {O'Shea},
  \citenamefont {Reiser},\ and\ \citenamefont {Walter}}]{Kishek2003}%
  \BibitemOpen
  \bibfield  {author} {\bibinfo {author} {\bibfnamefont {R.~A.}\ \bibnamefont
  {Kishek}}, \bibinfo {author} {\bibfnamefont {S.}~\bibnamefont {Bernal}},
  \bibinfo {author} {\bibfnamefont {C.~L.}\ \bibnamefont {Bohn}}, \bibinfo
  {author} {\bibfnamefont {D.}~\bibnamefont {Grote}}, \bibinfo {author}
  {\bibfnamefont {I.}~\bibnamefont {Haber}}, \bibinfo {author} {\bibfnamefont
  {H.}~\bibnamefont {Li}}, \bibinfo {author} {\bibfnamefont {P.~G.}\
  \bibnamefont {O'Shea}}, \bibinfo {author} {\bibfnamefont {M.}~\bibnamefont
  {Reiser}}, \ and\ \bibinfo {author} {\bibfnamefont {M.}~\bibnamefont
  {Walter}},\ }\href {\doibase 10.1063/1.1558291} {\bibfield  {journal}
  {\bibinfo  {journal} {Physics of Plasmas}\ }\textbf {\bibinfo {volume}
  {10}},\ \bibinfo {pages} {2016} (\bibinfo {year} {2003})}\BibitemShut
  {NoStop}%
\bibitem [{\citenamefont {Bernal}\ \emph {et~al.}(1999)\citenamefont {Bernal},
  \citenamefont {Kishek}, \citenamefont {Reiser},\ and\ \citenamefont
  {Haber}}]{Bernal1999}%
  \BibitemOpen
  \bibfield  {author} {\bibinfo {author} {\bibfnamefont {S.}~\bibnamefont
  {Bernal}}, \bibinfo {author} {\bibfnamefont {R.}~\bibnamefont {Kishek}},
  \bibinfo {author} {\bibfnamefont {M.}~\bibnamefont {Reiser}}, \ and\ \bibinfo
  {author} {\bibfnamefont {I.}~\bibnamefont {Haber}},\ }\href {\doibase
  10.1103/PhysRevLett.82.4002} {\bibfield  {journal} {\bibinfo  {journal}
  {Physical Review Letters}\ }\textbf {\bibinfo {volume} {82}},\ \bibinfo
  {pages} {4002} (\bibinfo {year} {1999})}\BibitemShut {NoStop}%
\bibitem [{\citenamefont {Shishlo}\ \emph {et~al.}(2015)\citenamefont
  {Shishlo}, \citenamefont {Cousineau}, \citenamefont {Holmes},\ and\
  \citenamefont {Gorlov}}]{Shishlo2015}%
  \BibitemOpen
  \bibfield  {author} {\bibinfo {author} {\bibfnamefont {A.}~\bibnamefont
  {Shishlo}}, \bibinfo {author} {\bibfnamefont {S.}~\bibnamefont {Cousineau}},
  \bibinfo {author} {\bibfnamefont {J.}~\bibnamefont {Holmes}}, \ and\ \bibinfo
  {author} {\bibfnamefont {T.}~\bibnamefont {Gorlov}},\ }\href {\doibase
  10.1016/j.procs.2015.05.312} {\bibfield  {journal} {\bibinfo  {journal}
  {Procedia Computer Science}\ }\textbf {\bibinfo {volume} {51}},\ \bibinfo
  {pages} {1272} (\bibinfo {year} {2015})}\BibitemShut {NoStop}%
\bibitem [{\citenamefont {Ruisard}\ \emph {et~al.}(2020)\citenamefont
  {Ruisard}, \citenamefont {Aleksandrov}, \citenamefont {Cousineau},
  \citenamefont {Shishlo}, \citenamefont {Tzoganis},\ and\ \citenamefont
  {Zhukov}}]{Ruisard2020a}%
  \BibitemOpen
  \bibfield  {author} {\bibinfo {author} {\bibfnamefont {K.}~\bibnamefont
  {Ruisard}}, \bibinfo {author} {\bibfnamefont {A.}~\bibnamefont
  {Aleksandrov}}, \bibinfo {author} {\bibfnamefont {S.}~\bibnamefont
  {Cousineau}}, \bibinfo {author} {\bibfnamefont {A.}~\bibnamefont {Shishlo}},
  \bibinfo {author} {\bibfnamefont {V.}~\bibnamefont {Tzoganis}}, \ and\
  \bibinfo {author} {\bibfnamefont {A.}~\bibnamefont {Zhukov}},\ }\href@noop {}
  {\bibfield  {journal} {\bibinfo  {journal} {in preparation}\ } (\bibinfo
  {year} {2020})}\BibitemShut {NoStop}%
\bibitem [{\citenamefont {Crandall}\ and\ \citenamefont
  {Wangler}(1988)}]{Crandall1988a}%
  \BibitemOpen
  \bibfield  {author} {\bibinfo {author} {\bibfnamefont {K.~R.}\ \bibnamefont
  {Crandall}}\ and\ \bibinfo {author} {\bibfnamefont {T.~P.}\ \bibnamefont
  {Wangler}},\ }\href@noop {} {\bibfield  {journal} {\bibinfo  {journal} {AIP
  Conference Proceedings}\ }\textbf {\bibinfo {volume} {177}},\ \bibinfo
  {pages} {22} (\bibinfo {year} {1988})}\BibitemShut {NoStop}%
\bibitem [{\citenamefont {Henderson}\ \emph {et~al.}(2014)\citenamefont
  {Henderson}, \citenamefont {Abraham}, \citenamefont {Aleksandrov},
  \citenamefont {Allen}, \citenamefont {Alonso}, \citenamefont {Anderson},
  \citenamefont {Arenius}, \citenamefont {Arthur}, \citenamefont {Assadi},
  \citenamefont {Ayers}, \citenamefont {Bach}, \citenamefont {Badea},
  \citenamefont {Battle}, \citenamefont {Beebe-Wang}, \citenamefont
  {Bergmann},\ and\ \citenamefont {Others}}]{Henderson2014}%
  \BibitemOpen
  \bibfield  {author} {\bibinfo {author} {\bibfnamefont {S.}~\bibnamefont
  {Henderson}}, \bibinfo {author} {\bibfnamefont {W.}~\bibnamefont {Abraham}},
  \bibinfo {author} {\bibfnamefont {A.}~\bibnamefont {Aleksandrov}}, \bibinfo
  {author} {\bibfnamefont {C.}~\bibnamefont {Allen}}, \bibinfo {author}
  {\bibfnamefont {J.}~\bibnamefont {Alonso}}, \bibinfo {author} {\bibfnamefont
  {D.}~\bibnamefont {Anderson}}, \bibinfo {author} {\bibfnamefont
  {D.}~\bibnamefont {Arenius}}, \bibinfo {author} {\bibfnamefont
  {T.}~\bibnamefont {Arthur}}, \bibinfo {author} {\bibfnamefont
  {S.}~\bibnamefont {Assadi}}, \bibinfo {author} {\bibfnamefont
  {J.}~\bibnamefont {Ayers}}, \bibinfo {author} {\bibfnamefont
  {P.}~\bibnamefont {Bach}}, \bibinfo {author} {\bibfnamefont {V.}~\bibnamefont
  {Badea}}, \bibinfo {author} {\bibfnamefont {R.}~\bibnamefont {Battle}},
  \bibinfo {author} {\bibfnamefont {J.}~\bibnamefont {Beebe-Wang}}, \bibinfo
  {author} {\bibfnamefont {B.}~\bibnamefont {Bergmann}}, \ and\ \bibinfo
  {author} {\bibnamefont {Others}},\ }\href {\doibase
  10.1016/j.nima.2014.03.067} {\bibfield  {journal} {\bibinfo  {journal}
  {Nuclear Instruments and Methods in Physics Research, Section A:
  Accelerators, Spectrometers, Detectors and Associated Equipment}\ }\textbf
  {\bibinfo {volume} {763}} (\bibinfo {year} {2014}),\
  10.1016/j.nima.2014.03.067}\BibitemShut {NoStop}%
\bibitem [{\citenamefont {Ratti}\ \emph {et~al.}(2000)\citenamefont {Ratti},
  \citenamefont {Digennaro}, \citenamefont {Gough}, \citenamefont {Hoff},
  \citenamefont {Keller}, \citenamefont {Kennedy}, \citenamefont {Macgill},
  \citenamefont {Staples}, \citenamefont {Virostek},\ and\ \citenamefont
  {Yourd}}]{Ratti2000}%
  \BibitemOpen
  \bibfield  {author} {\bibinfo {author} {\bibfnamefont {A.}~\bibnamefont
  {Ratti}}, \bibinfo {author} {\bibfnamefont {R.}~\bibnamefont {Digennaro}},
  \bibinfo {author} {\bibfnamefont {R.~A.}\ \bibnamefont {Gough}}, \bibinfo
  {author} {\bibfnamefont {M.}~\bibnamefont {Hoff}}, \bibinfo {author}
  {\bibfnamefont {R.}~\bibnamefont {Keller}}, \bibinfo {author} {\bibfnamefont
  {K.}~\bibnamefont {Kennedy}}, \bibinfo {author} {\bibfnamefont
  {R.}~\bibnamefont {Macgill}}, \bibinfo {author} {\bibfnamefont
  {J.}~\bibnamefont {Staples}}, \bibinfo {author} {\bibfnamefont
  {S.}~\bibnamefont {Virostek}}, \ and\ \bibinfo {author} {\bibnamefont
  {Yourd}},\ }in\ \href@noop {} {\emph {\bibinfo {booktitle} {Proceedings of
  EPAC 2000}}}\ (\bibinfo {address} {Vienna, Austria},\ \bibinfo {year}
  {2000})\ pp.\ \bibinfo {pages} {495--497}\BibitemShut {NoStop}%
\end{thebibliography}%

\end{document}